\documentclass[]{spie}  

\usepackage{amsmath,amsfonts,amssymb}
\usepackage{graphicx}
\usepackage[colorlinks=true, allcolors=blue]{hyperref}
\usepackage{nicematrix}

\title{Self-Supervised Equivariant Regularization Reconciles Multiple Instance Learning: Joint Referable Diabetic Retinopathy Classification and Lesion Segmentation}

\author[$a^{\dagger}$]{Wenhui Zhu}
\author[$b^{\dagger}$]{Peijie Qiu}
\author[$c$]{Natasha Lepore}
\author[d]{Oana M. Dumitrascu}
\author[a*]{Yalin Wang}
\affil[a]{School of Computing and Augmented Intelligence, Arizona State University, AZ 85281, USA;}
\affil[b]{McKeley School of Engineering, Washington University in St. Louis, St. Louis, MO 63130, USA;}
\affil[c]{CIBORG Lab, Department of Radiology Children’s Hospital Los Angeles, Los Angeles, CA 90027, USA;}
\affil[d]{Department of Neurology, Mayo Clinic, Scottsdale, AZ 85251, USA}

\authorinfo{$\dagger$ The authors contribute equally to this paper \\ *Send correspondence to Yalin Wang\\ E-mail: ylwang@asu.edu}

\begin{document} 
\maketitle

\begin{abstract}
Lesion appearance is a crucial clue for medical providers to distinguish referable diabetic retinopathy (rDR) from non-referable DR. Most existing large-scale DR datasets contain only image-level labels rather than pixel-based annotations. This motivates us to develop algorithms to classify rDR and segment lesions via image-level labels. This paper leverages self-supervised equivariant learning and attention-based multi-instance learning (MIL) to tackle this problem. MIL is an effective strategy to differentiate positive and negative instances, helping us discard background regions (negative instances) while localizing lesion regions (positive ones). However, MIL only provides coarse lesion localization and cannot distinguish lesions located across adjacent patches. Conversely, a self-supervised equivariant attention mechanism (SEAM) generates a segmentation-level class activation map (CAM) that can guide patch extraction of lesions more accurately. Our work aims at integrating both methods to improve rDR classification accuracy. We conduct extensive validation experiments on the Eyepacs dataset, achieving an area under the receiver operating characteristic curve (AU ROC) of 0.958, outperforming current state-of-the-art algorithms. 
\end{abstract}


\section{INTRODUCTION}
 The International Clinical Diabetic Retinopathy Disease Severity Scale grades the severity of diabetic retinopathy (DR) according to the characteristic lesion area, and separates them into the following classes: no retinopathy, mild non-proliferative DR (NPDR), moderate NPDR, severe NPDR, and proliferative DR (PDR) \cite{Wu2013ClassificationOD}. No retinopathy or Mild NPDR is defined as non-referable DR that has no obvious pathological features. On the contrary, the severity beyond moderate is defined as referable DR(rDR). Delayed diagnosis of referable diabetic retinopathy (rDR) may cause severe vision loss and is likely to result in blindness. As a result, an automatic DR diagnosis framework is crucial in clinical practice to help patients receive proper treatments, lowering the risk of severe vision damage. 
 
 The critical difference between referable and non-referable DR is the appearance of the various lesions, such as retinal haemorrhage or exudate, which are the main bio-markers to help ophthalmologists differentiate them. The semantic information of lesion, e.g.,  boundary and intensity, helps us localize and categorize different lesions and hence facilitate rDR grading. However, most of the existing large-scale DR datasets lack information on lesion regions, e.g., pixel-level annotations for rDR lesions.  Manually annotating lesion segmentation masks is a laborious process and requires medical expertise. To mimic how ophthalmologists diagnose rDR and take advantage of a large amount of weakly annotated (i.e., image-level annotations) retinal images, we propose a self-supervised equivariant regularization joint with a multiple instance learning (MIL) to classify rDR and segment lesion regions.
 
\begin{figure}[t]
\includegraphics[width=\textwidth]{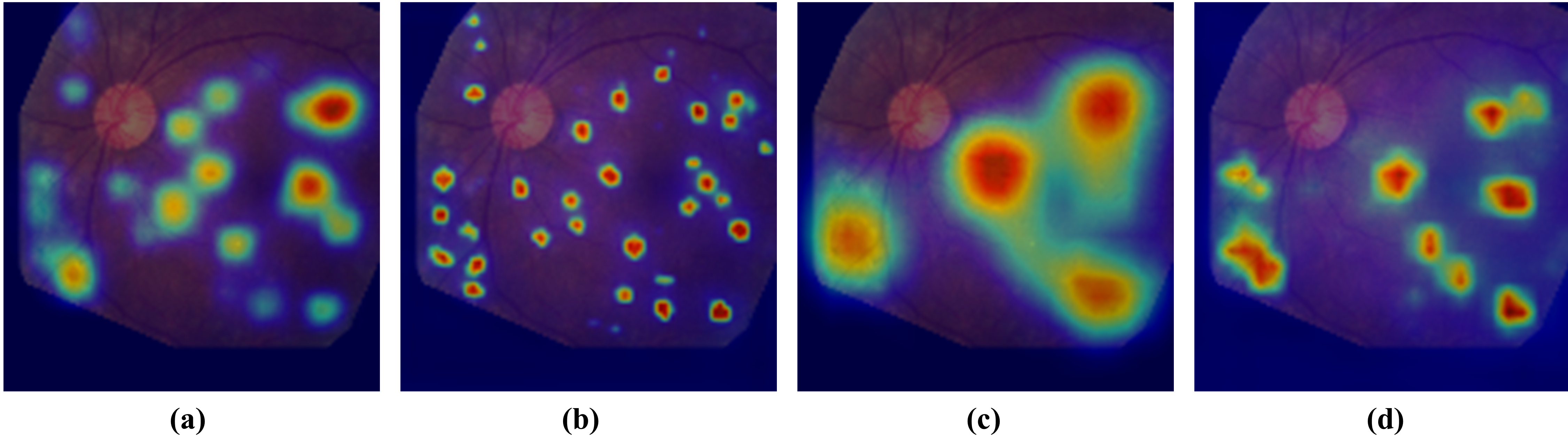}
\caption{The generated class activation maps (CAM) from the training monitoring process: (a) CAM from the original image branch in the Siamese network, (b) refined CAM from the original image branch, (c) CAM from the affine-transformed branch in the Siamese network, (d) refined CAM from the affine-transformed branch. There is a difference between the CAM of the original image and that of the affine-transformed image. The refined CAM produces a better localization boundary of lesion regions compared to the original CAM.} \label{fig1}
\end{figure}

In recent years, there has been significant progress in the diagnosis of rDR through deep learning-based methods. Most of this work to date has focused on rDR classification in supervised frameworks, taking advantage of different architectures. Some studies \cite{graham2015kaggle,quellec2017deep} focused on exploring the effects of different pre-processing techniques, including image pyramid, enhancement selection, and improved training on neural network performance. For instance, Huang et al.\cite{Huang2021IdentifyingTK} demonstrated that ResNet50 \cite{He2016DeepRL} achieved better classification performance after optimizing the pre-processing and training strategies. Attention-based multiple instance learning was used in \cite{Ilse2018AttentionbasedDM} to localize the patches within a retinal image that contribute most to the final prediction. However, the method only provided patch-level lesion localization, and each lesion was likely to cover only a small portion of a patch. It is challenging to differentiate positive and negative instances if a lesion region is divided into adjacent patches. Sadafi et al.\cite{Sadafi2020AttentionBM} mitigated this problem by using R-CNN to filter out potential instances and then feeding them into an attention-based MIL. Nevertheless, this method cannot provide segmentation-level localization of the lesions. Some weakly supervised segmentation methods were implemented \cite{Ahn2018LearningPS,Kolesnikov2016SeedEA,Fan2020CIANCA} that leveraged class activation mapping (CAM) \cite{Zhou2016LearningDF} to provide a segmentation mask. In addition, Wang et al.\cite{Wang2020SelfSupervisedEA} proposed a self-supervised equivariant attention mechanism (SEAM) to narrow the gap between classification and segmentation tasks via an equivariant regularization and self-attention-based CAM refinement.SEAM is a semantic segmentation framework, the main task lies in the refinement of lesions, and the segmentation results need to be further applied to the classification task to improve the accuracy.

In this paper, we propose a novel method based on self-supervised equivariant regularization and attention-based multiple instances learning to jointly classify rDR and produce lesion segmentations. As shown in Fig.~\ref{fig1}, the refined CAM from the SEAM mechanism can localize lesion regions accurately, which assists the MIL in accurately selecting positive and negative patches. Simultaneously, MIL fine-tunes the SEAM module to produce lesion-localized CAM. The three main contributions of our work can be summarized as: (1) We propose a self-supervised method that provides an implicit regularization to guide the MIL to select accurate positive and negative instances. (2) A framework to obtain segmentation-level CAM based on weakly-supervised image-level labels is introduced. (3) The MIL prediction module, based on an accurate lesion selection provided by SEAM, considerably improves rDR classification performance. To the best of our knowledge, this is the first work to simultaneously classify rDR while generating lesion segmentation with only image-level labels.

\section{Methods}
\begin{figure}[t]
\includegraphics[width=\textwidth]{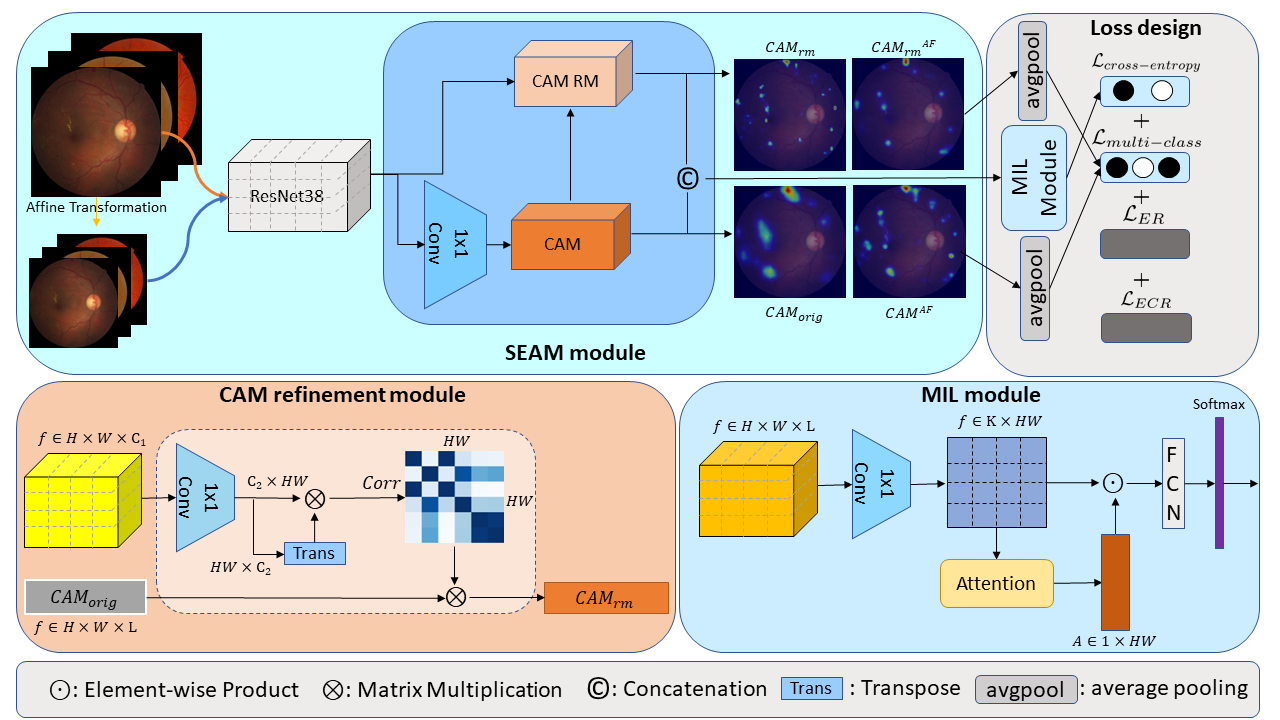}
\caption{The network architecture of our proposed method is implemented as a Siamese network where $CAM_{orig}, CAM_{rm}$  denotes the original CAM and refined CAM, respectively, from the original branch in the Siamese network, and $CAM^{AF}, CAM_{rm}^{AF}$ denotes the CAM and refined CAM from the affine-transformed branch, respectively.} 
\label{fig:network}
\vspace{-0.1cm}
\end{figure}

The proposed method consists of two main modules, a SEAM module, and a modified attention-based MIL one. The SEAM mechanism serves to localize the lesion areas and produces a fine-grained CAM as a byproduct, that approximates the ground truth lesion segmentation. The equivariant regularization narrows the gap between the classification and segmentation tasks by moving the CAM toward the ground truth segmentation of the sought-after object. The MIL module can further localize pathologically important regions for grading Diabetic Retinopathy within an image by representing an image with a bag of features.
Our modified attention-based MIL module formulates bags of instances directly from the feature space instead of feeding every single patch into the network, saving a lot of computation time compared to the conventional patch-based MIL. The rationale behind this is that differentiating positive and negative instances in the feature space is more effective than that in the patch-based MIL because equivariant regularization helps the feature space encode meaningful semantic information of lesions. Our proposed method is implemented as a Siamese network~\cite{Bromley1993SignatureVU} with ResNet38~\cite{Wu2019WiderOD} as the backbone; the network architecture is shown in Fig.~\ref{fig:network}.  

\subsection{Self-Supervised Equivariant Attention Module}

\noindent \textbf{Equivariant Regularization.}
A fundamental assumption of DL-based segmentation is that feeding an affine-transformed (e.g., rescale, rotation, flip, etc.) input image to a segmentation network results in approximately the same affine transformation in the segmentation output. In contrast, the pooling layer in  classification networks destroys this property~\cite{Ahn2018LearningPS}. Penalizing this inequivalence pushes the CAM of a target object toward its corresponding segmentation. The equivariant regularization (ER) loss can be given by
\begin{equation}
    \mathcal{L}_{ER} = || AF(CAM_{orig}) - CAM^{AF}||_{1},
\end{equation}

\noindent where $CAM_{orig}, \ CAM^{AF}$ denotes the CAM from the original image and the affine-transformed image, respectively, and $AF(\cdot)$ denotes the affine transformation applied to the original image. Using the $L_1$ norm in the ER loss ensures the sparsity of the CAM because lesions typically only account for a small portion of a retinal image.

\noindent \textbf{CAM Refinement.}
The CAM is further refined by the correlation between pixels, based on the fact that similar pixels ought to belong to the same object. The correlation is computed as cosine similarity between pixels as
\begin{equation}
    Corr(f_i, f_j) = ReLU(\frac{\theta(f_i)^{T} \theta(f_j)}{||\theta(f_i)|| \cdot ||\theta(f_j) ||}),
\end{equation}
where $f_i, f_j$ denotes the feature map at spatial location ${i, j}$, and $\theta$ is an embedding function to reduce the number of channels given by a $1\times 1$ convolution layer. The $ReLU$ activation is to avoid negative values in the correlation matrix. The refined CAM is then computed as:
\begin{equation}
    CAM_{rm}(f_i) = \frac{1}{N(f_i)} \sum_{j} Corr(f_i, f_j) \cdot CAM_{orig}(f_i),
\end{equation}
where $N(f_i) = \sum_{j}Corr(f_i, f_j) $ denotes a normalization constant.


\noindent \textbf{Equivariant Cross Regularization.}
In our experiments, we found that the affine-transformed branch in the Siamese network identifies more lesion areas than the original branch. The refined CAMs in each branch lose information, making the original ER loss fall into a local minimum where all pixels belong to a single class. This problem can be mitigated by cross-regulating the original CAM and refined CAM in both branches of the Siamese network because the CAM from those two branches focuses on different regions. The equivariant cross regularization (ECR) loss, which helps us reduce CAM degeneration and escape local minima, is given as 
\begin{equation}
     \mathcal{L}_{ECR} = || AF(CAM_{orig}) - CAM_{rm}^{AF}||_{1} + ||  AF(CAM_{rm})-CAM^{AF}||_{1},
\end{equation}
where $CAM_{rm}^{AF}$ denotes the refined CAM of the affine-transformed input image.

\subsection{Attention-Based Multiple Instance Learning Module}
We view the MIL learning task as solving two fundamental problems: how to extract instances and how to aggregate them. For the former problem, conventional patch-based instance extraction divides an image into a number $N$ of $p\times p$ patches, where each patch serves as an instance. In our early experiments, however, patch-based MIL~\cite{Ilse2018AttentionbasedDM} with attention pooling had difficulty in differentiating the positive and negative instances. The contribution weights between positive and negative are close to each other, particularly when: 1) the lesion region is extremely small, 2) a single lesion area spans across adjacent patches during patch selection. For the latter problem, different pooling strategies~\cite{Pinheiro2015FromIT,Zhu2017DeepMN,Feng2017DeepMN} have been proposed to aggregate instance-level representation. We use the attention pooling proposed by Ilse et al.~\cite{Ilse2018AttentionbasedDM} to assign a weight to each instance based on an attention mechanism.  

The feature map under self-supervised equivariant regularization encodes the global distribution of lesions. Instead of using the patch-based MIL method, we take advantage of the global lesion feature maps $f_{orig} \in R^{   L\times HW }$ and $f_{rv} \in R^{   L\times HW }$. Each element $f_{i,j}$ at the spatial location $i, j$ of the feature map $f$ is treated as an instance, resulting in $H\times W$ is the number of instances. We aggregate the information from both $f_{orig}$ and $f_{rv}$ together and increase the number of channels to $K$ by a $1 \times 1$ convolution layer to enrich the feature space, resulting in a final feature map $f \in R^{ K\times HW  }$. The MIL Attention mechanism, which weights the contribution of each instance, is given by
\begin{equation}
      A =Sigmoid \left({{ w^{\top }_{1} ReLU(w_{2}f^{\top })} } \right),
\end{equation}
where $w_{1}\in R^{ D\times 1 }$, $w_{2}\in R^{ D\times K}$, we apply the attention map A to the feature map $f \in R^{  K\times HW  }$, 
    \begin{equation}
     f_{MIL} = f\odot A
    \end{equation}
where $\odot$ denotes element-wise multiplication. The feature map is fed into a fully-connected layer with a softmax activation to output probabilistic predictions.

\subsection{Loss Design}
The network is jointly trained with self-supervised equivariant regularization and MIL, outputting classification prediction, and a fine-grained lesion CAM. The overall objective function is designed as 
\begin{equation}
    \mathcal{L} = \mathcal{L}_{multi-class} + \mathcal{L}_{ER} + \mathcal{L}_{ECR} +  \mathcal{L}_{cross-entropy}.
\end{equation}
The $\mathcal{L}_{cross-entropy}$ is the cross-entropy loss used for optimizing the output from MIL for classification prediction, given by 
\begin{equation}
    \mathcal{L}_{cross-entropy} = -\sum_{c=1}^{C}y_{c} \ log(p_{c}),
\end{equation}
where $y_{c} \in \{0, 1\}$ is a binary label to indicate whether it is rDR, and $p_c$ is predicted probability of class $c$. The $\mathcal{L}_{multi-class}$ is a multi-label soft margin loss, which adapts the cases that multiple labels exist for one image, and is used for optimizing the CAM, given by   
\begin{equation}
     \mathcal{L}_{multi-class} = -\frac{1}{C-1}\sum_{c=1}^{C-1} y_{c}^{'} \ log(\frac{1}{1+e^{-y_{o, c}}}) + (1-y_{c}^{'}) log(\frac{e^{-y_{o, c}}}{1+e^{-y_{o, c}}}).
\end{equation}
Here $C-1$ denotes the number of target objects (we only have one foreground  object-lesion-in our cases), $y_{c}^{'} \in \{0, 1\}$ denotes whether lesion exists, and $y
_{o, c}$ is the output predicted CAM after an adaptive-average pooling layer. Note that $y_{c}^{'}$ consists of both foreground and background labels, representing whether a target object or background exists, respectively. In our case, we one-hot encoded the foreground and background regions for each image. For example, the \{1, 1\} represents referable DR  with background, and \{1, 0\} represents No-referable DR label with background, where the first label indicates whether background exists, and the second indicates whether lesion exists.

\begin{figure}[t]
\includegraphics[width=\textwidth]{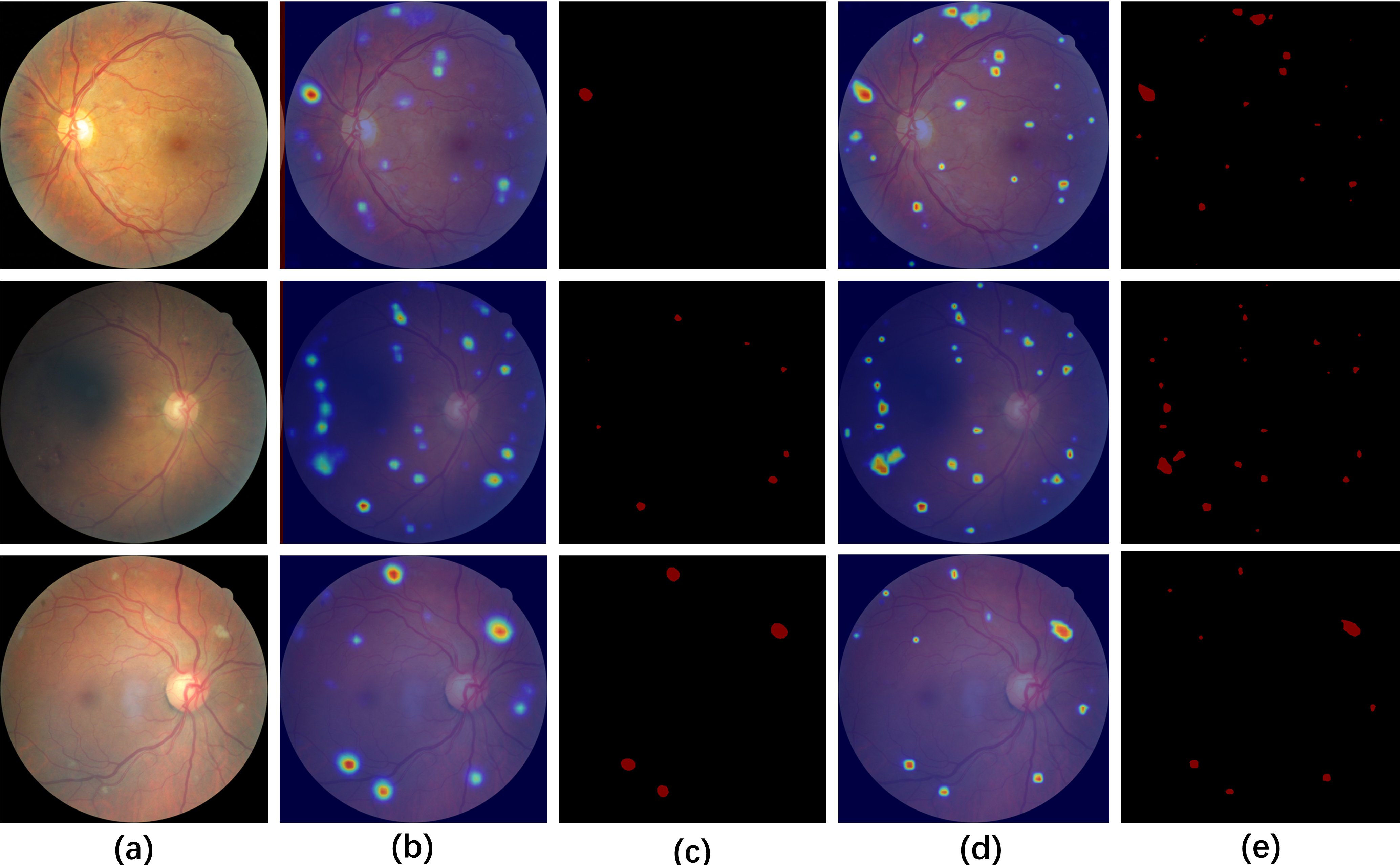}
\caption{(a) Original input image, (b) original CAM, (c) lesion segmentation from original CAM, (d) refined CAM, (e) lesion segmentation from refined CAM. The segmentation generated by the original CAM is likely to lose lesion information. Our proposed method provides better localization in terms of lesion boundary and preserves more information about lesion regions.}
\label{fig:refinedCAM}
\end{figure}

\section{Experiments}
We validate our proposed method on the Eyepacs dataset \cite{cuadros2009eyepacs}, through three experiments: (i) ResNet 38, (ii) attention-based MIL with ResNet38 as backbone, and (iii) our proposed self-supervised equivariant regularized attention based MIL with ResNet38 as backbone. We also compare the performance of our model with other state-of-the-art work. 

\noindent \textbf{Dataset.}
The Eyepacs dataset consists of 35,126 training images, 10,906 validation images, and 42,670 testing images. The original dataset consists of 4 levels grading of DR: mild, moderate, severe non-proliferative, and proliferative. In clinical practice, DR grading beyond the mild stage is recognized as referable DR; otherwise, it is non-referable DR. All images are center-cropped and resized to size of $512 \times 512$. 

\noindent \textbf{Data Augmentation.}
To prevent overfitting and increase generalizability, massive data augmentation is performed, including random horizontal flips, vertical flips, random crops, extra color jitters, random rotations, and random translations. 
Note that the same data augmentation is deployed in all our experiments. 

\noindent \textbf{Implementation Details.}
ResNet38~\cite{Wu2019WiderOD}, excluding the global average pooling layer and final fully connected layer, is used as the backbone for all our experiments, because of its state-of-the-art performance on semantic segmentation. A rescaling transformation with a factor of 0.4 is deployed as our affine transformation when training our proposed method. All models were trained with a stochastic gradient descent (SGD) optimizer, but we assigned different learning rates to different parameter groups: lower learning rates to the parameters of ResNet38, and larger ones to others. The initial learning rate is set to 0.001, with a polynomial decay with a rate of 0.9. A weight decay with rate of 0.0005 is applied to further prevent overfitting. All our experiments were performed on an Nvidia GeForce RTX 2080 with a batch size of 3 and a number of training epochs of 100. 


\noindent \textbf{Evaluation Metrics.}
 We evaluated the rDR classification via accuracy, F1 score, and area under the receiver operating characteristic curve (AU ROC).

\subsection{Baseline Experiment}
The baseline experiment is based on the split of the training set, validation set, and testing set of the official EyePacs dataset.  It evaluates the importance of different modules by comparing the accuracy, F1 score, and AU ROC. As shown in Table~\ref{tab:CompOtherbaseline}, comparative experiments are conducted on Resnet38, the MIL module, and the SEAM module, respectively, with the same training configuration. All the experimental results are reported in Table 1 for rDR classification in terms of accuracy, F1 score, and AU ROC. The highest AU ROC score of our proposed method is \textbf{0.958}. The original CAM, refined CAM, and segmentation was generated on the test set to visually inspect and validate our proposed method, as shown in Fig.~\ref{fig:refinedCAM}.

\begin{table}[htp]
\centering
\caption{Comparison of our proposed method (ResNet38+SEAM + MIL) with other baselines.
Mean and standard deviation for accuracy, F1 score, and area under ROC curve (AU ROC). \vspace{0.1cm} }\label{tab:CompOtherbaseline}
\begin{NiceTabular}{*{1}{p{5.0cm}}*{4}{c} }[hvlines]
\centering \textbf{Method} & \textbf{Dataset} & \textbf{Accuracy} & \textbf{F1 Score} & \textbf{AU ROC}  \\ 
\Block{2-1}{ResNet38} & Val set & 0.931  & 0.809  & 0.954  \\ & Test set & 0.928  & 0.805  & 0.950  \\ 
\Block{2-1}{ResNet38 +  MIL (Patch-based)} & Val set & 0.936  & 0.818  & 0.957  \\ & Test set  & 0.932  & 0.810  & 0.953  \\ \Block{2-1}{\textbf{ResNet38 + SEAM + MIL{(Ours)}}} & Val set & \textbf{0.939}  & \textbf{0.829}  & \textbf{0.961}  \\ & Test set  & \textbf{0.936}  & \textbf{0.823}  & \textbf{0.958}  \\ 
\end{NiceTabular}
\label{table2}
\end{table}

\subsection{Comparison Experiments}
We compared our results with some state-of-the-art work in terms of AU ROC in Table~\ref{tab:SOTA}. Our proposed method outperforms the listed state-of-the-art work. Note that we used the official split training set with only data augmentation to train our model, and evaluated it on the official split validation and test datasets, while other methods included a series of preprocessing and training strategies as detailed in \cite{rakhlin2018diabetic,graham2015kaggle,quellec2017deep, Pires2019ADA}. 


\begin{table}[ht]
\centering
\caption{Comparison of our proposed method with other state-of-the-art methods on the Eyepacs dataset for rDR classification. \vspace{0.1cm} }\label{tab:SOTA}
\begin{tabular}{
  |p{\dimexpr.4\linewidth-2\tabcolsep-1.3333\arrayrulewidth}
  |p{\dimexpr.35\linewidth-2\tabcolsep-1.3333\arrayrulewidth}
  |p{\dimexpr.25\linewidth-2\tabcolsep-1.3333\arrayrulewidth}|
 }

\hline
\textbf{Dataset Split} & \textbf{Method} &  \textbf{AU ROC} \\
\hline

  Training(all) \ \  Testing(all)&Leibig et al. \cite{leibig2017leveraging}    & 0.927            \\  
  &Rakhlin et al. \cite{rakhlin2018diabetic}   & 0.920              \\ 
  &Pires et al. \cite{Pires2019ADA}     & 0.946           \\  
  &Graham et al. \cite{graham2015kaggle}     & 0.951            \\  
  &Quellec et al. \cite{quellec2017deep}    & 0.954         \\ \hline
  
   Training(all) \ \ Testing(all)& \textbf{MIL-SEAM(Ours)}    &\textbf{0.958}         \\  \hline
\end{tabular}
\end{table}

\section{Conclusion}

In this paper, we proposed to regularize multiple instances learning to bag more accurately positive and negative instances for rDR classification, using a self-supervised equivariant regularization mechanism. As a byproduct, we obtained a fine-grained CAM and a coarse segmentation of lesions with weakly supervised image-level labels, guiding medical providers to diagnose rDR. Our ablation study on the Eyepacs dataset showed improvement of our proposed method over the patch-based attention-based MIL method. The performance of our proposed method is also comparable to the state-of-the-art work, achieving a 0.958 AU ROC.

The application of our proposed method can be extended to many other medical imaging applications for which only image-level annotations are available, and where a specific series of patterns are directly related to predictive models. We provided an example of adapting SEAM to multiple instances learning to mitigate its drawbacks to our task. At the same time, predictive models other than multiple instance learning can be adjusted based on different tasks.

\bibliography{main} 
\bibliographystyle{spiebib} 

\end{document}